# Fabry–Pérot resonances and a crossover to the quantum Hall regime in ballistic graphene quantum point contacts

Nurul Fariha Ahmad[1,2], Katsuyoshi Komatsu[1], Takuya Iwasaki[3], Kenji Watanabe[4], Takashi Taniguchi[4], Hiroshi Mizuta[5,6], Yutaka Wakayama[1], Abdul Manaf Hashim[2], Yoshifumi Morita[7], Satoshi Moriyama[1], Shu Nakaharai[1†]

[1]International Center for Materials Nanoarchitectonics (WPI-MANA), National Institute for Materials Science (NIMS), Tsukuba, Ibaraki 305-0044, Japan.

[2]Malaysia-Japan International Institute of Technology, Universiti Teknologi Malaysia, Jalan Sultan Yahya Petra, 54100 Kuala Lumpur, Malaysia.

[3]International Center for Young Scientists (ICYS), National Institute for Materials Science (NIMS), Tsukuba, Ibaraki 305-0044, Japan.

[4]Research Center for Functional Materials, NIMS, Tsukuba, Ibaraki 305-0044, Japan.

[5]School of Material Science, Japan Advanced Institute of Science and Technology, Nomi, Ishikawa 923-1211, Japan.

[6]Hitachi Cambridge Laboratory, Hitachi Europe Ltd., J. J. Thomson Avenue, Cambridge, United Kingdom.

[7]Faculty of Engineering, Gunma University, Kiryu, Gunma 376-8515, Japan.

† Corresponding author. Email: NAKAHARAI.Shu@nims.go.jp

**Abstract**

We report on the observation of quantum transport and interference in a graphene device that is attached with a pair of split gates to form an electrostatically-defined quantum point contact (QPC). In the low magnetic field regime, the resistance exhibited Fabry–Pérot (FP) resonances due to *np'n*(*pn'p*) cavities formed by the top gate. In the quantum Hall (QH) regime with a high magnetic field, the edge states governed the phenomena, presenting a unique condition where the edge channels of electrons and holes along a *p–n* junction acted as a solid-state analogue of a monochromatic light beam. We observed a crossover from the FP to QH regimes in ballistic graphene QPC under a magnetic field with varying temperatures. In particular, the collapse of the QH effect was elucidated as the magnetic field was decreased. Our high-mobility graphene device enabled observation of such quantum coherence effects up to several tens of kelvins. The presented device could serve as one of the key elements in future electronic quantum optic devices.

**Main**

In conventional ballistic 2D electron gases such as a heterojunction of GaAs, constriction due to split-gate electrodes, referred to as quantum point contacts (QPC), creates a conducting channel by electrostatic confinement. On the other hand, QPCs in graphene cannot directly create such electrostatically confined channels due to the absence of a bandgap in this material; despite the absence of quantised transport at low magnetic fields, quantum interference phenomena show intriguing behaviour in graphene QPCs. There are alternative methods to create such structures in graphene, e.g. etched constrictions [1,2] or narrow and long electrostatic channels [3]. Moreover, graphene has a favourable feature of ballistic conduction with an extremely long mean free path, which can be further improved by sandwiching the material between hexagonal boron nitride (hBN) sheets.

Much attention has been paid recently to electronic quantum optics, in which the fabrication of a collimated electron interferometer plays a central role [4], and whose scope ranges from Veselago lensing to fundamental studies of negative refraction [5,6], analogous to electromagnetic-wave scattering in negative index materials [7]. The structure where the cone-shaped energy bands of electrons and holes meet at their apexes, called the Dirac point, is also a unique characteristic of graphene 2D electron system, producing a distinctive series of Landau levels in the QH regime. As electrons are confined in two dimensions, they move along the edge of the device under a strong magnetic field, which is referred to as QH edge channel. These channels are a solid-state analogue of a monochromatic light beam. The QH edge states show striking quantum interference effects as they are protected against backscattering. In particular, $p$–$n$ junctions in graphene have the potential to provide a basic building block for electronic quantum optics [8, 9, 10, 11, 12]. Figures 1(a,b) show the QPC structure most suitable for manipulating such confined channels. The graphene QPC offers a unique mechanism for electron transport in the QH regime with gate-defined constrictions, where edge modes travel along the boundaries of locally-gated regions [13, 14]. The narrowly separated $p$–$n$ interfaces of the QPC are controlled between open and closed configurations by the top-gate voltage ($V_{TG}$), and this structure could serve as a crucial component in future graphene electronic quantum optics devices. The electronic states in graphene are easily accessible by local gates. Furthermore, the encapsulation of graphene with hBN makes the fabrication process more tractable compared with previous studies.

In this study, we report on ballistic electron transport through a QPC formed in high-mobility graphene via electrostatic control by a pair of split gates. As detailed in Methods below, the graphene is encapsulated by hBN, resulting in hBN/Graphene/hBN stacks. The thickness of the used top and bottom hBN is 25 nm and 34 nm, respectively. As the magnetic field $B$ was increased, we observed a crossover from the Fabry–Pérot (FP) regime to the QH regime. In the FP regime, the conductance exhibited resonances due to

*np'n*(*pn'p*) cavities formed by the top gate, while in the QH regime, the edge states governed the transport phenomena and presented a unique condition where edge channels of electrons and holes along a *p–n* junction play a central role. Crossover between FP and QH regimes has not been explored in ballistic graphene QPCs. Because our graphene sample is attached to an atomically flat and inert hBN surface, the effects of the surface roughness and impurities in the substrate are greatly reduced, leading to high mobility of up to 200,000 cm$^2$/Vs and ballistic transport. Our device furthers the pathway towards quantum coherent phenomena in graphene devices, such as fractional QH states and shot noise experiments.

The longitudinal resistance ($R_\mathrm{L}$) across the QPC (Fig. 1(c,d)), as a function of $V_\mathrm{TG}$ and back-gate voltage ($V_\mathrm{BG}$), demonstrates the independent carrier control on the bulk and top-gated regions at $T = 6$ K in the absence of magnetic field ($B = 0$ T), the mapping of which is shown in Fig. 2(a). Carrier density in the top-gated regions is controlled by both $V_\mathrm{TG}$ and $V_\mathrm{BG}$, while that in the bulk region (away from the top gates) is controlled solely by $V_\mathrm{BG}$. The $V_\mathrm{TG}$–$V_\mathrm{BG}$ mapping of $R_\mathrm{L}$ in Fig. 2(a) shows the presence of *np'n*(*pn'p*) and *nn'n*(*pp'p*) regimes, and their boundaries, indicated by broken red lines, correspond to the charge neutrality condition; the horizontal line represents the charge neutrality of the bulk regions, while the inclined line is of the top-gated regions. Along the inclined line, charges generated by the top gates are negated by those of the back gates, i.e. $C_\mathrm{TG}V_\mathrm{TG} + C_\mathrm{BG}V_\mathrm{BG} = 0$; therefore, the slope of the inclined line is determined by the ratio of the capacitances of both gates. Here we obtain $C_\mathrm{TG} = 7.6 \times 10^{-8}$ F/cm$^2$ and $C_\mathrm{BG} = 1.9 \times 10^{-8}$ F/cm$^2$ deduced from the Hall analysis of the back gate. At the intersection of these two lines ($V_\mathrm{TG}$, $V_\mathrm{BG}$; ~0 V, 0.2 V), the entire sample is in a charge neutral state. Under a strong magnetic field, graphene should be in the QH regime and the $V_\mathrm{TG}$–$V_\mathrm{BG}$ mapping is altered by Landau quantisation, as shown in Fig. 2(c), at $T = 6$ K and $B = 2$ T. The FP regime, QH regime and a crossover from FP- to-QH- regime are discussed further in the subsequent texts.

*FP regime:* As shown in Fig. 2(b), $R_\mathrm{L}$ in *np'n*(*pn'p*) regimes exhibits oscillations, which also appear in the mapping of $dR_\mathrm{L}/dV_\mathrm{TG}$. The top-gated regions of these regimes serve as a resonant cavity for the coherent electron trajectory, while the pair of *p–n* junctions acts as semi-reflective mirrors. The oscillation of the resistance value indicates quantum interference effects bouncing between two *p–n* junctions in the resonant cavity of the top-gated region [15, 16]. Although the QPC structure overlaps with the resonant cavity, FP resonance is still observed, guaranteeing a collimating effect on ballistically transmitted carriers (see also Supplementary Information for detail of the characteristic scales). In this low magnetic field regime, the QPC structure is irrelevant and the device behaves as two *p–n* junctions. In terms of the magnetic-field dependence of the FP resonances, FP reflects a salient feature of the Berry phase in graphene [17] that is closely related to the Klein tunnelling physics [15, 16, 18]. The magnetic-field dependence of the resistance pattern in the low magnetic-field regime is shown in Fig. 3(a) for a $V_{BG}$ of 1.2 V. As the magnetic field increases, the overall pattern shifts gradually towards the negative $V_{TG}$ side and shows a discontinuous half-a-period phase shift when $B\sim50$ mT, which is consistent with previous studies on top-gated graphene devices. Moreover, the phase shift implies the presence of Klein tunnelling, as it is a signature of the perfect transmission of carriers normally occurring on the junctions.

Figure 4 shows the temperature dependence of $R_L$ in which the resistance oscillations weaken as the temperature rises; oscillations remain at 20 K but fade out above 40 K. The high-mobility graphene device in this study makes it possible to observe this quantum coherence effect at such high temperatures. This crossover condition is consistent with the picture at which thermal fluctuations are comparable to the phase difference between interfering paths in the device, taking into account the effective width of the top-gated region with a length of approximately 200 nm (see also Supplementary Information Fig. S1(b)) [16].

*QH regime:* In the QH regime of the QPC structure with a high magnetic field, the edge states govern the phenomena as originally discussed in ref. [13]. Here, we present the updated data by employing the hBN/Graphene/hBN device with high mobility. Moreover, the effect of gap width is analysed recently in detail in ref. [19]. As shown in Fig. 2(c), the QH effects and the edge-mode mixing under a finite magnetic field govern transport properties. The configuration of edge modes in the vicinity of the QPC is switched between open- and closed- states by gate voltage control [13]. In the 'open' configuration, $R_L$ vanishes because the edge modes do not contribute to the total resistance. In contrast, $R_L$ has a finite value and is quantised in a 'closed' QPC configuration as some edge modes are backscattered. Both top and back gates can manipulate the open/closed state of edge modes. The assignment of the quantised conductance in our hBN/graphene/hBN QPC device is discussed in ref. [19].

*FP to QH crossover:* As the magnetic field increases, a crossover from the FP to QH regimes takes place. In the beginning, we focus on the bipolar *np'n(pn'p)* regimes as shown in Fig.3 (b). In the semi-classical picture, the electron trajectory bends under a magnetic field due to the Lorentz force. When the electron cyclotron radius under the top-gated regions becomes comparable to the cavity size, the basic set-up undergoes a crossover to the edge-state-transport/QH regime. In the crossover, 'snake trajectories' play a crucial role [20, 21, 22, 23]. The fade out of FP oscillation is set by the magnetic field $\hbar k_F/eL_{cav}$, where $L_{cav}$ is an effective cavity length. In reference to the Supplementary Information, the typical $L_{cav}$ of our device is ~200 nm, with a characteristic magnetic field of ~0.2 T in the carrier density we focused on, which gives a consistent picture. Under high magnetic field conditions, Shubnikov–de Haas (SdH) oscillations take place due to the formation of Landau levels in both unipolar and bipolar regimes, indicating that the FP oscillations are connected to the SdH oscillations of the QH regime. Therefore, as the magnetic field is increased, the QPC structure governs the quantised charge transport through itself rather than through the *np'n(pn'p)* cavity. In this case, the QH regime evolves owing to the confinement to the edge channels along the *p–n* junction electrostatically formed by the top gates. SdH periodic oscillations in the QH regime are superimposed by the parabola-shaped lines, the envelope of which is indicated by broken yellow lines in Fig. 3(b). As this characteristic feature is not observed in rectangular top gate geometries [16], it is deemed to be induced by the QPC structure. We consider this as closely related to the 'snake' oscillations pointed out in the *p-n* junction context [20,21,22,23], in which a sequence of half of the closed cyclotron motions in the *p* and *n* regions forms the snake trajectory. Along the snake trajectory, the carrier is either transmitted or backscattered through the QPC, which can lead to the oscillatory behaviour with a parabola-shaped envelope, as depicted in Fig. 3(b). This phenomenon can be quantitatively described in future studies.

Figure 3(c) shows the unipolar or *nn'n(pp'p)* regimes, which convey distinct oscillation behaviour. While FP resonances evolve into SdH signals in the bipolar regime, the SdH signal bends in the unipolar regime of *nn'n* and *pp'p* configurations and the oscillation is strongly suppressed as the magnetic field decreases [24].

To summarise, we have demonstrated a quantum transport at cryogenic conditions from a zero-magnetic field regime to a QH regime in a graphene device that is attached with a pair of split gates, to form an electrostatically-defined QPC. In the bipolar regime of *np'n* and *pn'p*, FP periodic oscillations appear in the longitudinal differential resistance across the QPC, as the carrier density of the *np'n* (*pn'p*) cavity increases under weak magnetic fields. As the magnetic field increased FP oscillations evolved into SdH oscillations. Furthermore, novel oscillation phenomena not previously observed in a rectangular top-gate case, emerge during the evolution of the oscillations from the FP to QH regimes, attributable to the QPC structure. Although we consider these oscillations to be caused by 'snake states' transmitted or backscattered through the QPC, details of this phenomenon are yet to be solved. The present knowledge shall lead to a more comprehensive characterisation of the collapse of the QH effect and the crossover between the FP and QH regimes.

Note added: During the preparation of the present paper for submission, we noticed that a very recent preprint [25] reports the overlapped complementary topics.

**Methods**

Graphene and hBN layers were prepared using a mechanical exfoliation method, and the stacked layers of hBN/graphene/hBN were fabricated via an all-dry transfer method [26,27,28]. The resulting structure of a single-layer graphene sandwiched by hBN was placed on Si wafer with a 90-nm-thick silicon dioxide ($SiO_2$) layer. The thickness of the top and bottom hBN is 25 nm and 34 nm, respectively. Subsequently, samples were patterned into the Hall bar geometry by e-beam lithography and etched via reactive ion etching. One-dimensional edge contacts were formed by the deposition of metals (Cr/Au) on the edge of the stack [27], while a split-gate structure was simultaneously formed with a 60-nm gap located at the centre of the Hall bar (Fig. 1(b)). The QPC device was measured using a four-terminal configuration under the standard lock-in measurement technique. $R_L$ was estimated by measuring the voltage difference between terminals 2 and 3 when a fixed amplitude current of 1–10 nA was applied at a frequency of 17 Hz between the source (terminal 1) and drain (terminal 4) (Fig. 1(b)). The Dirac point was detected at nearly zero back-gate voltage (~0.2 V) with a very narrow and sharp resistance peak, which indicated the presence of both high quality and very low intrinsically-doped graphene. High mobilities of 120,000 $cm^2$/Vs (electron) and 200,000 $cm^2$/Vs (hole) were confirmed with an estimated mean free path >1 µm, comparable to the device geometry, which implied ballistic transport. Low temperature measurements were performed down to 6 K under perpendicular magnetic fields up to 6 T. Landau quantisation of bulk graphene was presented as a Landau fan (Fig. 1(c,d)), indicating a high quality of the hBN/graphene/hBN device for both electron and hole configurations. In addition, the degeneration of Landau levels was demonstrated. In this study, no satellites of the main Dirac point were observed within accessible gate voltage owing to misalignment between graphene and hBN, in contrast to the results of a previous study [28].

**Acknowledgements**

The device fabrication and measurement were supported by the Japan Society for Promotion of Science (JSPS) KAKENHI 26630139; and the NIMS Nanofabrication Platform Project, the World Premier International Research Center Initiative on Materials Nanoarchitectonics, sponsored by the Ministry of Education, Culture, Sports, Science and Technology (MEXT), Japan. Growth of hBN was supported by the Elemental Strategy Initiative conducted by the MEXT, Japan and JSPS KAKENHI 15K21722, and the CREST (JPMJCR15F3), JST. N.F. Ahmad thanks NIMS and UTM for the financial support during her research attachment at NIMS and Malaysia Ministry of Education for the scholarship.

**Author Contributions**

S.N. and S.M. conceived and designed the experiments. N.F.A., and K.K., fabricated the devices. N.F.A. and S.M. performed the experiments. T.I., H.M., and Y.M. performed the simulations. N.F.A., T.I., Y.M. S.M. and S.N. analyzed the data, developed the models, and wrote the paper. All authors discussed and edited the manuscript. K.W. and T.T. provided the hBN crystals used in the devices.

## Competing financial interests

The authors declare no competing interests.

## Data and materials availability

All data needed to evaluate the conclusions in the paper are present in the paper and/or the Supplementary Materials. Additional data related to this paper may be requested from the authors.

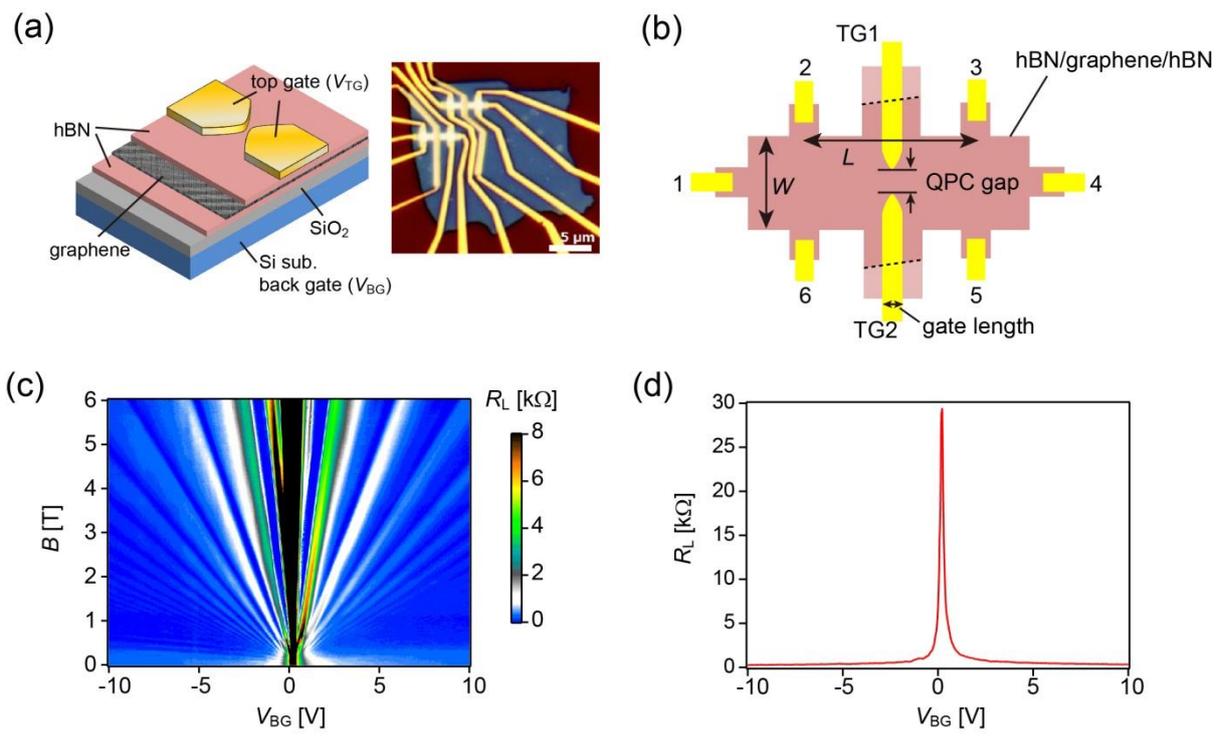

Fig. 1

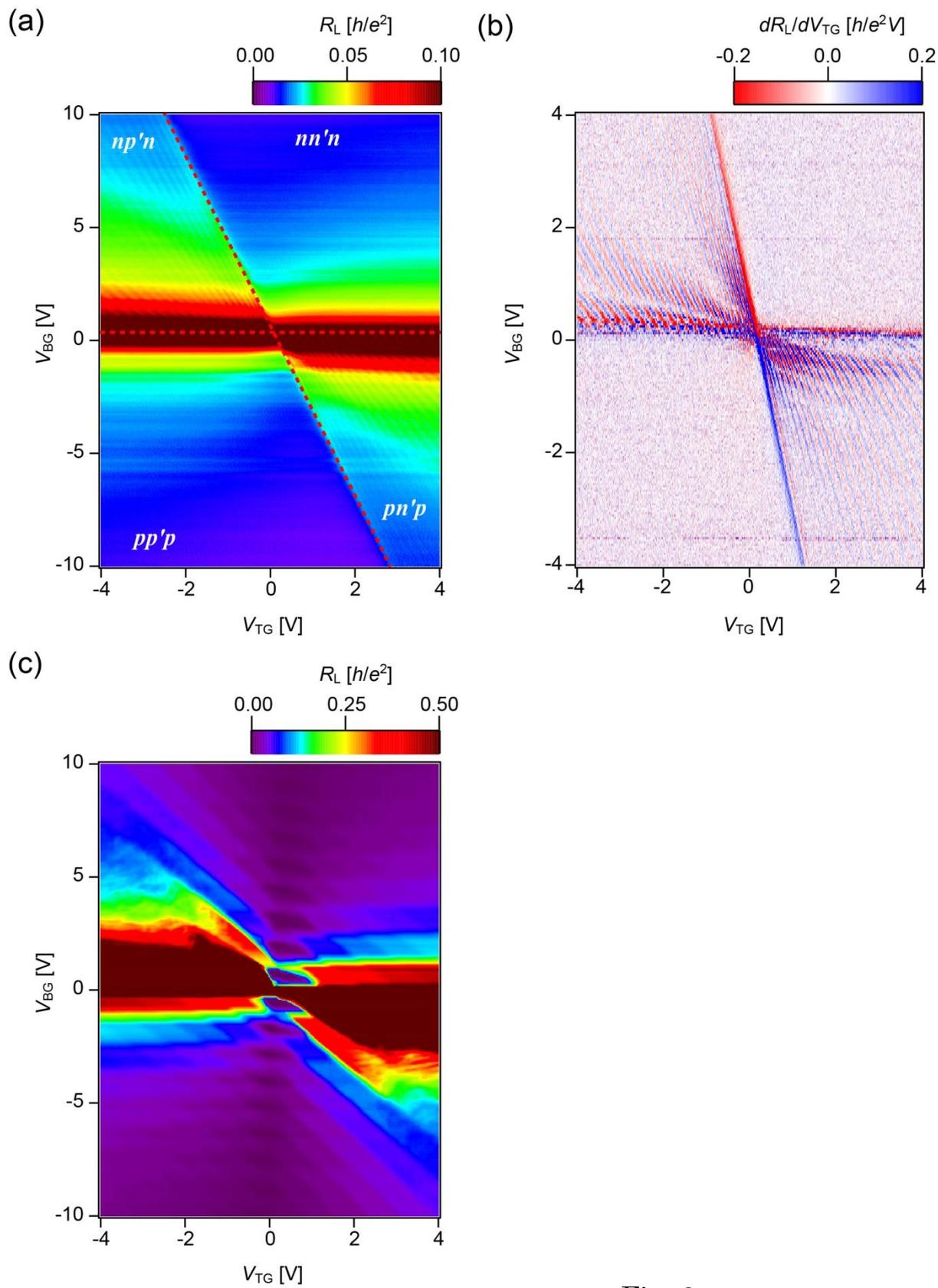

Fig. 2

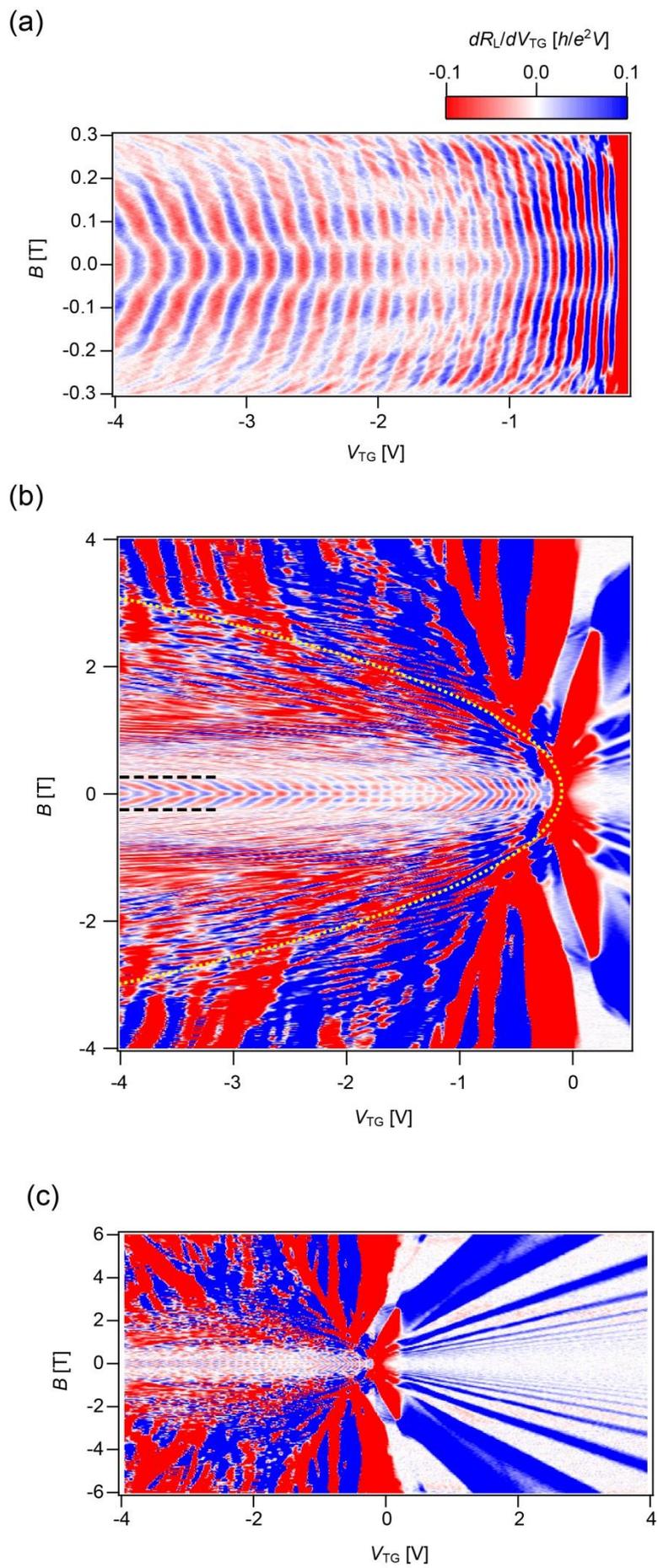

Fig. 3

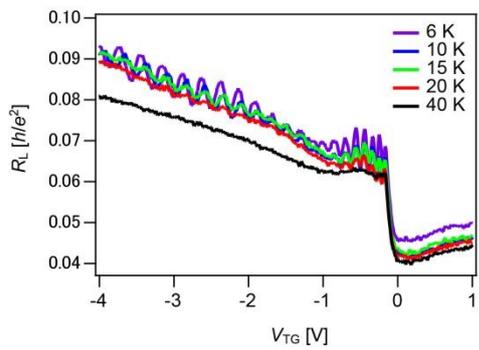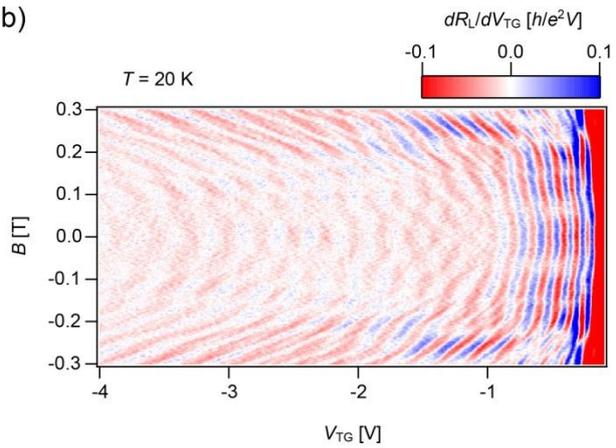

Fig. 4

**Figure Captions**:

**Figure 1:** Device structure and characterisation of the studied hBN/graphene/hBN QPC. **(a)** A schematic (left) and optical image (right) of the fabricated device. **(b)** Hall bar geometry of the device with dimensions of $L = 3.2$ μm, $W = 0.8$ μm, gate length of 150 nm and QPC gap of 60 nm. Six contacts 1–6 are attached to the graphene using edge contacts. Top gates are also attached, where broken lines indicate the edge of the graphene. **(c)** Landau fan diagram, mapping of $R_L$ as a function of $V_{BG}$ and $B$, at $V_{TG} = 0$ V and $T = 6$ K. **(d)** Longitudinal resistance $R_L$ vs. back-gate voltage $V_{BG}$, which corresponds to $B = 0$ T in (c).

**Figure 2:** $V_{TG}$–$V_{BG}$ mapping of the resistance. **(a)** $V_{TG}$–$V_{BG}$ mapping of $R_L$ at $B = 0$ T and $T = 6$ K. Horizontal and inclined broken lines indicate the charge neutrality conditions of the bulk and top-gated regions, respectively. **(b)** $V_{TG}$–$V_{BG}$ mapping of $R_L$ differentiated by $V_{TG}$ under the same conditions as in (a). FP oscillations are visible in *np'n* and *pn'p* regimes. **(c)** $V_{TG}$–$V_{BG}$ mapping of $R_L$ under magnetic field $B = 2$ T, which is in the QH regime.

**Figure 3:** Magnetic field ($B$) evolution of the resistance ($R_L$) differentiated by top-gate voltage ($V_{TG}$) as a function of $V_{TG}$. The magnetic-field dependence of the resistance pattern in the low magnetic-field regime is shown at $V_{BG} = 1.2$ V. **(a)** Gate-voltage and magnetic-field dependence of the oscillatory conductance of the FP regime in low magnetic field. As magnetic field is increased, the overall pattern gradually shifts towards the negative top-gated side and shows a discontinuous half-a-period phase shift at around $B = 50$ mT, implying Klein tunnelling effects. **(b)** In higher magnetic fields, FP oscillations evolve into SdH oscillations. Parabolic lines (yellow broken lines), described by an envelope of additional oscillations, are superimposed on the figure as guide for the eye. **(c)** Mapping for full range of top-gate voltage $V_{TG}$ swept from –4 V to 4 V.

**Figure 4:** Temperature dependence of the resistance ($R_L$) as a function of top-gate voltage ($V_{TG}$). **(a)** Temperature dependence of FP oscillations (see also Fig. 3) from 6 K to 40 K in the *np'n* regime. Resistance oscillations weaken as temperature rises and vanishes above 40 K. **(b)** FP oscillations (see also Fig.3) are still visible at $T = 20$ K.

## Supplementary Information

**This PDF file includes:**

・FIG. S1. Electrostatic-potential simulation of our QPC device
・FIG. S2. Simulation of the Fabry–Pérot resonances

・note S1. Modeling of the Electrostatic Potential
・note S2. Simulation of the Fabry–Pérot resonances

S1. Modeling of the Electrostatic Potential

We performed a simulation of the electrostatic-potential based on the method of ref. [16, 24], which lead to our picture of cavity formation by the top gate. Figure S1(a) shows the schematic of our device for the simulation. We employed parameters for the model to simulate the experimental devices. The model consisted of a quantum point contact (QPC) geometry with a gap of 60 nm covered by a vacuum box. The simulation was aided by COMSOL Multiphysics (COMSOL Inc.), a modelling and calculating software based on the finite element method. During the simulation, we applied voltage to the split top gates and grounded the graphene layer, obtained the electric field ($E_z$) in the $z$-direction at the surface of graphene in a self-consistent manner and calculated the Fermi energy ($E$) of graphene using the formula, $E = \text{sgn}(-E_z) \times \hbar v_F \sqrt{\pi |E_z| \varepsilon_0 \varepsilon_{hBN}/e}$, where $\hbar$ is the reduced Planck constant, $v_F \sim 10^6$ m/s is the Fermi velocity of graphene, $e$ is the elementary charge and $\varepsilon_0$ and $\varepsilon_{hBN}$ = 3.9 are the vacuum permittivity and the dielectric constant of hexagonal boron nitride, respectively.

Figure S1(b) shows the electrostatic-potential mapping at the surface of the graphene layer where the origin (0,0) corresponds to the centre of the QPC. Figure S1(c) shows the energy profile along with the $x$-direction at $y$ = 200 nm (across the split top-gate). The energy under the top gate increased as the top-gate voltage was increased, indicating the formation of the cavity for Fabry–Pérot resonances.

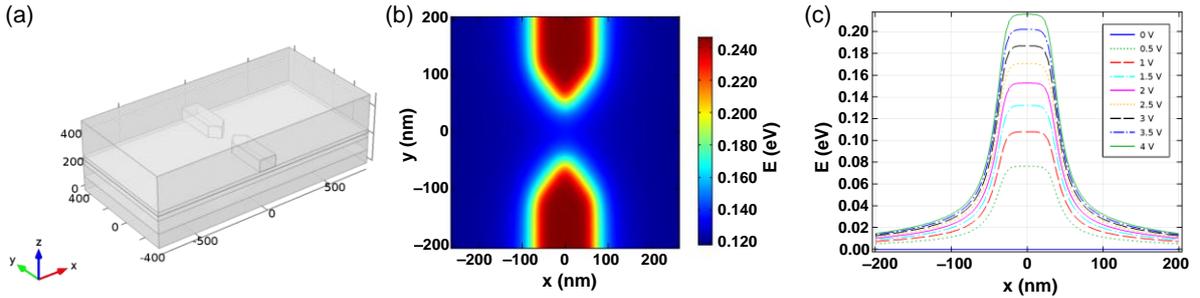

**FIG. S1:** Electrostatic-potential simulation of our QPC device. (a) Schematic illustration of the QPC device model (length unit: nm). (b) Mapping of the Fermi energy of graphene at the top surface for application of 4 V top-gate voltage. (c) Energy profile along the $x$-direction across the region under the top-gate ($y$ = 200 nm) for various top-gate voltages.

S2. Simulation of the Fabry–Pérot resonances

To substantiate our picture and fix characteristic energy scales, we performed a simulation of the transmission in our device through QPC. We applied WKB analysis based on the cavity model established in [15, 16]. Our focus was on clear FP resonances around $|V_{TG}|>\sim 1$V (Figs. 3 and 4 in our manuscript) where the half-a-period('$\pi$') phase shift takes place under a finite magnetic field, i.e. a signature of Klein tunnelling. A typical solution for the WKB analysis is given in Fig. S2 in a normalised unit and discussed below, while the oscillating part was focused on as in [16]. Results in S1 confirmed the basic characteristic scales [15]. Applying a typical value of an order 1 eV/$\mu$m$^2$ for the curvature of the potential ($=a$), we found a characteristic energy unit scale $(av_F^2\hbar^2)^{1/3}(=\varepsilon_*)$ of ~10 meV. Considering the conversion factor from the top-gate voltage ~1/25, the basic energy scale for FP resonance was consistent (Fig. S2). Furthermore, we determined a typical unit scale for the magnetic field ($\Phi_0/2\pi(\hbar v_F/\varepsilon_*)^2$) of ~0.1 T ($\Phi_0$: unit magnetic flux), which is compatible with the magnetic field for the $\pi$-shift. Although this model was simplified, it gave a basic unit for the energy/magnetic field scales of the experimental results, combined with the data in S1. On the other hand, the resonance amplitude should be suppressed due to e.g. roughness in the cavity surface and finite temperature effects. For a discussion of the fine structures of the quantum interference pattern, more details of the QPC structure and its electrostatic profile will be provided in a separate paper.

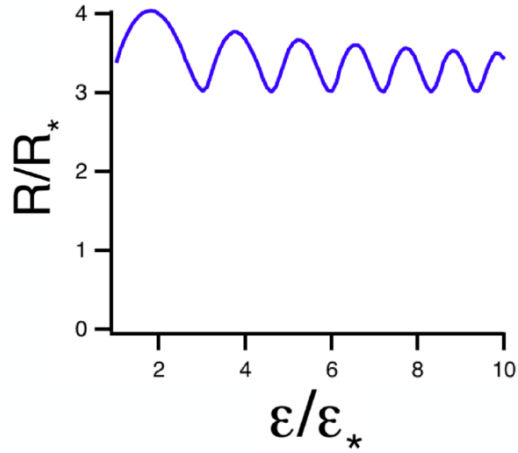

**FIG. S2:** Simulation of the Fabry–Pérot resonances. Typical resistance oscillation (in units of quantum resistance/scaled by a normalised cavity width) is shown as a function of energy at $B = 0$ T. Our focus is on the oscillatory behaviour and the offset is shifted for clarity. The unit scales are discussed above.